\documentclass[conference]{IEEEtran}
\usepackage{cite}
\usepackage[usenames,dvipsnames,svgnames,table]{xcolor}
\ifCLASSINFOpdf
   \usepackage[pdftex]{graphicx}
\else
\fi
\usepackage{amsmath}
\DeclareMathOperator*{\POPCOUNT}{POPCOUNT}
\usepackage{array}
\usepackage{siunitx}
\newcommand{\Vc}{\ensuremath{V_{\mathrm{c}}}}
\newcommand{\Rp}{\ensuremath{R_{\mathrm{P}}}}
\newcommand{\Rap}{\ensuremath{R_{\mathrm{AP}}}}
\newcommand{\muo}{\ensuremath{\mu_0}}

\begin{document}
\title{ Implementing Binarized Neural Networks with Magnetoresistive RAM without Error Correction}
\author{\IEEEauthorblockN{
Tifenn Hirtzlin\IEEEauthorrefmark{1}, 
Bogdan Penkovsky\IEEEauthorrefmark{1}, 
Jacques-Olivier Klein\IEEEauthorrefmark{1}, 
Nicolas Locatelli\IEEEauthorrefmark{1}, \\
Adrien F. Vincent\IEEEauthorrefmark{3},
Marc Bocquet\IEEEauthorrefmark{2}, 
Jean-Michel Portal\IEEEauthorrefmark{2} 
and 
Damien Querlioz\IEEEauthorrefmark{1}}
\IEEEauthorblockA{\IEEEauthorrefmark{1}Centre de Nanosciences et de Nanotechnologies, CNRS, Univ Paris-Sud, Universit\'e Paris-Saclay, 91129 Palaiseau, France\\ 
Email: damien.querlioz@u-psud.fr}
\IEEEauthorblockA{\IEEEauthorrefmark{2}Institut Mat\'eriaux Micro\'electronique Nanosciences de Provence, Univ. Aix-Marseille et Toulon, CNRS, France}
\IEEEauthorblockA{\IEEEauthorrefmark{3}Laboratoire de l'Int\'egration du Mat\'eriau au Syst\`eme, Univ. Bordeaux, Bordeaux INP, CNRS, Talence, France.}
}
\IEEEoverridecommandlockouts
\maketitle
\begin{abstract}
One of the most exciting applications of Spin Torque Magnetoresistive Random Access Memory (ST-MRAM) is the in-memory implementation of deep neural networks, which could allow improving the energy efficiency of Artificial Intelligence by orders of magnitude with regards to its implementation on computers and graphics cards. In particular,  ST-MRAM could be ideal for implementing Binarized Neural Networks (BNNs), a type of deep neural networks discovered in 2016, which can achieve state-of-the-art performance with a highly reduced memory footprint with regards to conventional artificial intelligence approaches. The challenge of ST-MRAM, however, is that it is prone to write errors and usually requires the use of error correction. In this work, we show  that these bit errors   can be tolerated by BNNs  to an outstanding level, based on examples of image recognition tasks (MNIST, CIFAR-10 and ImageNet): bit error rates of ST-MRAM up to  0.1\% have little impact on recognition accuracy. The requirements for ST-MRAM are therefore considerably relaxed for BNNs with regards to traditional applications. By consequence, we show that for BNNs, ST-MRAMs can be programmed with weak (low-energy) programming conditions, without error correcting codes. We show that this result can allow the use of low energy and low area ST-MRAM cells, and show that the energy savings at the system level can reach a factor two.
\end{abstract}


%
\IEEEpeerreviewmaketitle

\section{Introduction}

Spin Torque Magnetoresistive Random Access Memory (ST-MRAM) is currently emerging as a leading  technology for embedded memory in microcontroller units \cite{golonzka2018mram,lee2018embedded,shih2019logic}, as well as for standalone memory \cite{andre2017st}.
ST-MRAM indeed provides a compact fast and non volatile memory, which is fully embeddable in modern CMOS, and features outstanding endurance.
This unique set of features makes ST-MRAM also adapted for alternative non von Neumann computing schemes, which tightly integrate logic and memory \cite{zhao2014synchronous}.
In particular, this approach can be especially adapted for implementing hardware deep neural networks.
These algorithms have become the flagship approach of modern artificial intelligence  \cite{lecun2015deep}, but they possess a high power consumption, which is mainly caused by the von Neumann bottleneck \cite{editorial_big_2018}.    

Multiple proposals have been made to implement deep neural networks with  ST-MRAM using concepts of in-memory or near-memory computing \cite{natsui2018design,locatelli2018use,pan2018multilevel,he2018accelerating,ramasubramanian2014spindle}.
Similar ideas have been proposed for other types of resistive memory \cite{editorial_big_2018,ielmini2018memory,yu2018neuro,querlioz2015bioinspired}.
The benefit of ST-MRAM is its outstanding endurance \cite{golonzka2018mram,lee2018embedded}.
However, such proposals come with an important challenge:
ST-MRAMs feature bit errors. 
Commercial processes typically target a bit error rate (BER) of $10^{-6}$ \cite{golonzka2018mram,lee2018embedded}.
Such memories are therefore meant to be used  with error correcting codes, as is the case of other types of resistive memories \cite{ly2018role,ielmini2018memory}.
Unfortunately, the bit errors in ST-MRAMs are to a large extent intrinsic, as they can originate in the basic physics of spin torque-based magnetization switching  \cite{diao_spin-transfer_2007,vincent_analytical_2015}.
They will thus not disappear, even when the technology progresses.

In this paper, we look at the special case of Binarized Neural Networks (BNNs) \cite{courbariaux2016binarized,rastegari2016xnor}.
Theses networks have been proposed recently as a highly simplified form of deep neural networks, as both their neurons and synapses assume binary values during inference.
They therefore function with reduced memory requirements with regards to standard neural networks, and use extremely simple arithmetic operations (no multiplication).
They can nevertheless approach state-of-the-art performance on vision tasks on datasets such as CIFAR-10 or ImageNet \cite{courbariaux2016binarized,rastegari2016xnor,lin2017towards}.
Due to their simplicity and reduced resource requirements, BNNs are particularly adapted to in-memory hardware implementation   \cite{yu2018neuro,bocquet2018}.
In this work, we look at bit error rate impact on BNNs designed with ST-MRAMs. 

This works extends our prior work on the implementation of BNNs with hafnium oxide-based Resistive RAM \cite{bocquet2018,hirtzlin2019outstanding} to the case of ST-MRAMs by taking into account the particular mechanism of intrinsic BER in these devices.
Additionally, our prior work relied on analysis of relatively simple machine learning tasks (MNIST and CIFAR-10), here we add the analysis of  a much more realistic task (ImageNet). 

The paper makes the following contributions: 
\begin{itemize}
    \item We simulate BNNs incorporating bit errors on the weights, and show that BERs up to $10^{-3}$ can be tolerated.  For the first time, the resilience of BNNs is demonstrated on the large scale ImageNet image recognition task. We deduce that ST-MRAMs can be used directly without ECC  (section~\ref{sec:error_tolerance}).
    \item We highlight that due to this extreme tolerance we can even reduce the programming energy of ST-MRAMs with regards to conventional applications. Based on a physical analysis, we show that a factor two in programming energy may be saved without impact on BNN accuracy (section~\ref{sec:energy_saving}).
\end{itemize}


\section{Binarized Neural Networks Can Tolerate ST-MRAM Bit Errors without ECC}
\label{sec:error_tolerance}

Binarized Neural Networks are a class of neural networks, in which the synaptic weights and the neuronal states, can only take two values: $+1$ or $-1$ (whereas they assume real values in conventional neural networks).
Therefore, the equation for neuronal activation $A$ in a conventional neural network
\begin{equation}
\label{eq:activ_real}
    A =  f \left( \sum_i W_iX_i \right),
\end{equation}
($X_i$ are  inputs of the neuron,  $W_i$ the synaptic weights and $f$ the neuronal activation function) simplifies into
\begin{equation}
\label{eq:activ_BNN}
    A = \mathrm{sign} \left( \POPCOUNT_i \left( XNOR \left( W_i,X_i \right) \right)-T \right).
\end{equation}
$T$ is the threshold of the neuron and is learned. $\POPCOUNT$  counts the number of $1s$, and $\mathrm{sign}$ is the sign function.

It should be noted that the synapses also feature real weights during training. 
The binarized weights, which are equal to the sign of the  real weights, are used in the equations of forward and backward passes, but the real weights are updated as a result of the learning rule  \cite{courbariaux2016binarized}.
As the real weights can be forgotten once the learning process is finished, BNNs are outstanding candidates for hardware implementation of neural network inference:

\begin{itemize}
    \item Area and energy expensive multiplications  in eq.~\eqref{eq:activ_real} are replaced by one-bit exclusive NOR (XNOR) operations.
    \item Neurons and synapses require a single bit of memory.
\end{itemize}

ASIC implementations of BNNs have been proposed using purely CMOS solutions \cite{ando2017brein}.
Nevertheless, the most attractive implementations propose using RRAMs or ST-MRAMs for implementing the synaptic weights, in architectures that tightly integrate logic and memory \cite{sun2018fully,sun2018xnor,tang2017binary,yu2018neuro,bocquet2018, natsui2018design,pan2018multilevel,he2018accelerating}.

However, a major challenge of RRAMs and ST-MRAMs is the presence of bit errors.
In the case of ST-MRAM, this issue is traditionally solved by using  ECC \cite{golonzka2018mram,lee2018embedded,shih2019logic} or special programming strategies \cite{lakys2012self}.
In this work, we look at the possibility to use ST-MRAMs directly to store the binarized synaptic weights, without relying on any of these techniques.

\begin{figure}[htbp]
	\centering
	\includegraphics[width=\columnwidth]{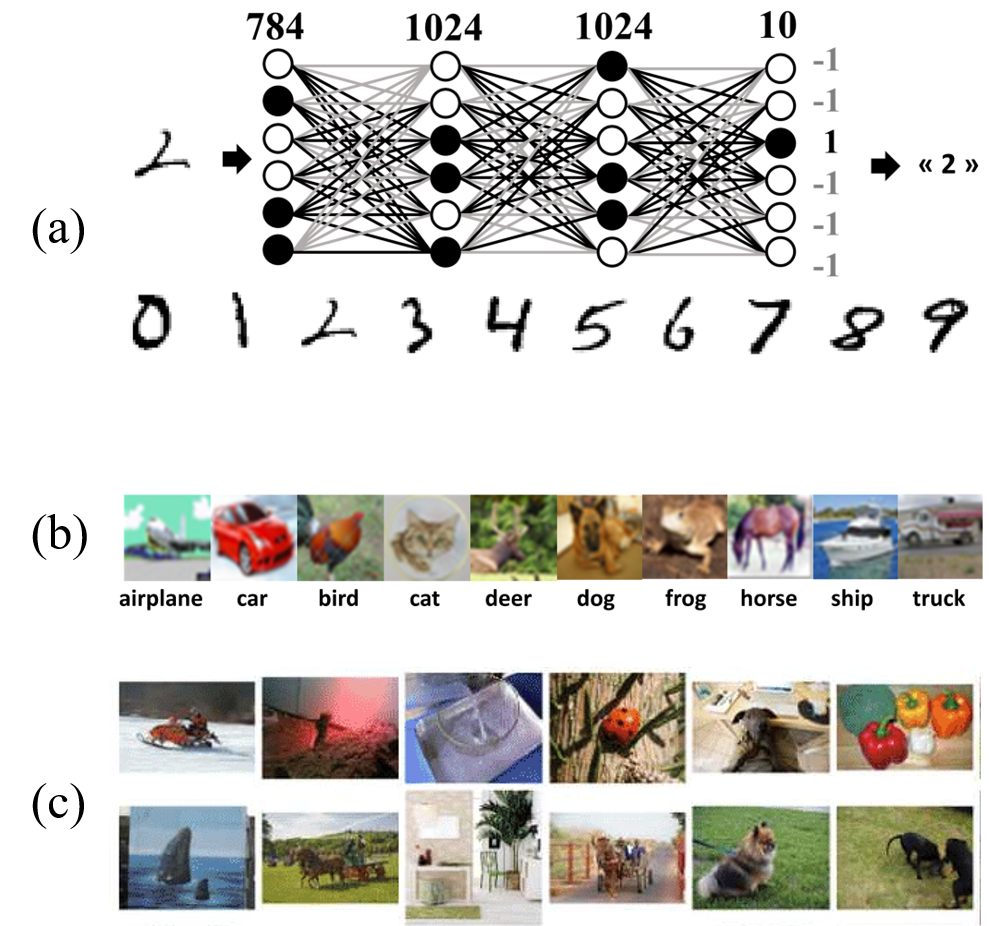}
	\caption{(a) Fully connected neural network used for the MNIST task, and example of MNIST dataset images. (b) Examples of CIFAR-10 dataset images. (c) Examples of ImageNet dataset examples.}
	\label{fig:BNN}
\end{figure}

\begin{figure}[htbp]
	\centering
	\includegraphics[width=\columnwidth]{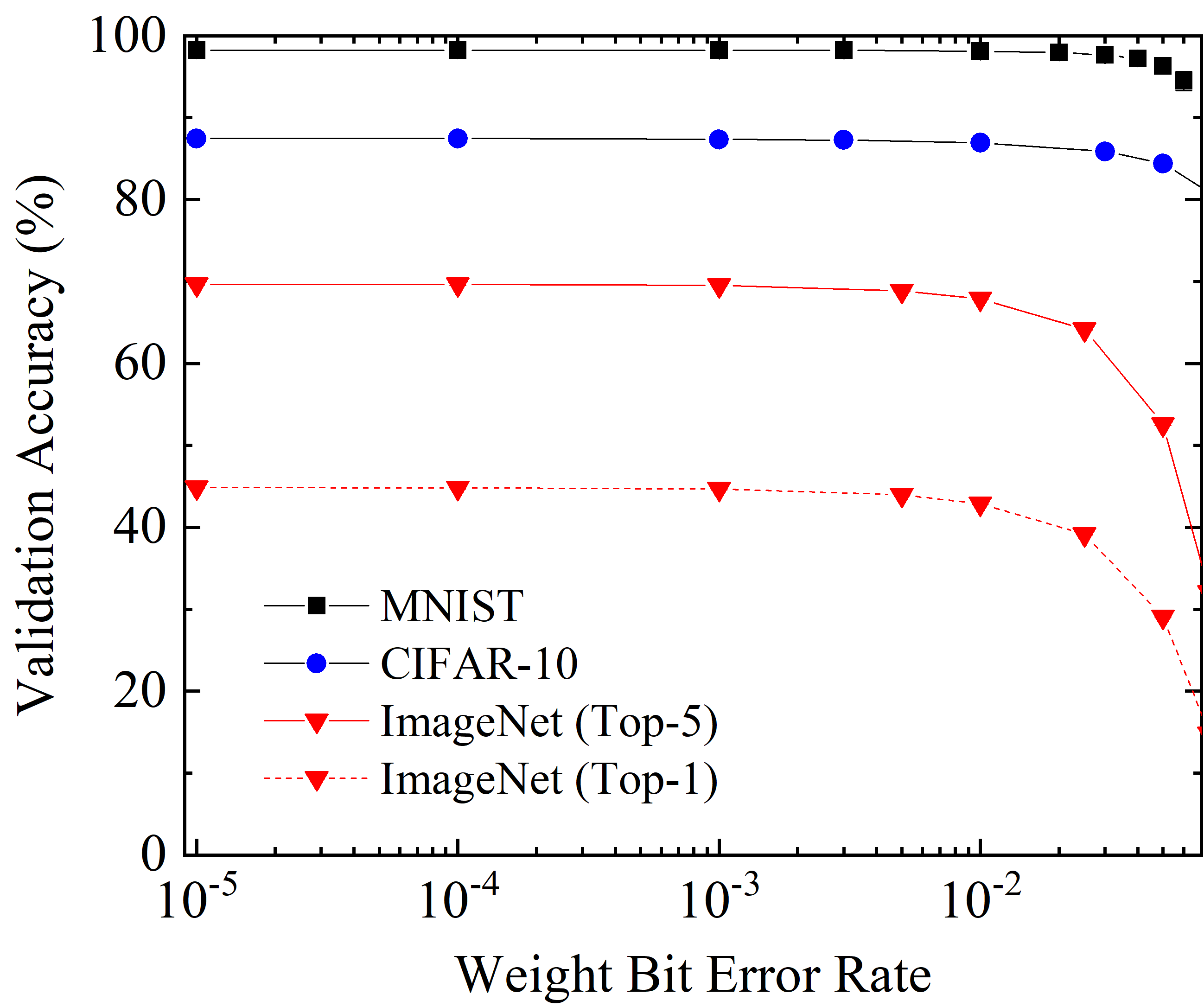}
		\caption{Recognition rate on the validation dataset of the fully connected neural network for MNIST, the convolutional neural network for CIFAR10, and AlexNet for ImageNet (Top-5 and Top-1) accuracies as a function of the bit error rate over the weights during inference.  
	    Each experiment was repeated five times, the mean recognition rate is presented.}
	\label{fig:accuracy}
\end{figure}

For this purpose, we perform simulations of BNNs, with bit errors added artificially.
We consider three type of neural networks (Fig.~\ref{fig:BNN}):
\begin{itemize}
    \item A shallow fully connected neural network with two 1024 neurons hidden layers, trained on the canonical task of MNIST handwritten digit classification.
    \item A deep convolutional network trained on the CIFAR-10 image recognition task. CIFAR-10 consists in  recognizing $32 \times 32$ color images out of ten classes of various animals and vehicles. Our neural network features 6 convolutional layers with number of filters 384, 384, 384, 768, 768 and 1536. 
    All convolutional layers use a  kernel size  of  $3 \times 3$, and a stride of one. These layers are topped by three fully connected layers with 1024, 1024 and 10 neurons.  
    \item The classic AlexNet deep convolutional neural network trained on the ImageNet recognition task \cite{krizhevsky2012imagenet}. ImageNet consists in  recognizing $128 \times 128$ color images out of 1000 classes, and is a considerably harder task than CIFAR-10. 
\end{itemize}

Training of the MNIST neural network is done with Python using NumPy libraries. 
For CIFAR-10, training is done using the TensorFlow framework.
For ImageNet, we use pretrained weights \cite{codeImagenet} obtained with the Caffe framework.
In all cases, a softmax function and cross-entropy loss are used during training.
Adam optimizer was used for better convergence, and dropout was used in the MNIST and CIFAR-10 cases to mitigate overfitting.
For inference, softmax is no longer needed and replaced by a hardmax function.

Fig.~\ref{fig:accuracy} shows the results of these simulations.
It plots the validation accuracy obtained for the three networks, as a function of the weight BER.
Each simulation was repeated five time, and the results were averaged.
In the case of ImageNet, we present both the Top-5 and Top-1 accuracies.

Interestingly, at BER up to $10^{-4}$, no effect on neural network accuracy is seen: the network performs just as well as when no error is present.
As ST-MRAMs can achieve BERs of $10^{-6}$,  \cite{golonzka2018mram,lee2018embedded}, this shows that they can be used directly, without ECC, for image recognition tasks. 
BERs of $10^{-3}$ also have practically no effect (the Top-5 accuracy on ImageNet is reduced from $69.7\%$ to  $69.5\%$). At a BER of $0.01$, the performance starts to decrease significantly.
The more difficult the task, the most substantial the performance reduction: MNIST accuracy is reduced from $98.3\%$ to $98.1\%$, CIFAR-10 accuracy is reduced from $87.65\%$ to $86.9\%$, ImageNet Top-5 accuracy is reduced from  $69.7\%$ to $67.9\%$.
These results highlight the inherent error correction capability of neural network, which  originates in their highly redundant nature.
It should be noted that these results could be further enhanced if one knows in advance the BER at inference time, by incorporating weight errors during training \cite{hirtzlin2019outstanding}.


\section{Saving Energy by Allowing More ST-MRAM Bit Errors}
\label{sec:energy_saving}

The results from section~\ref{sec:error_tolerance} suggest that ST-MRAMs can be used in a BNN with a BER much higher than the usually targeted $10^{-6}$, without error correction.
We now investigate the benefit that increasing the BERs can have in terms of energy consumption, based on physical modeling of ST-MRAM cells.

\begin{figure}[ht]
	\centering
	\includegraphics[width=2.0 in]{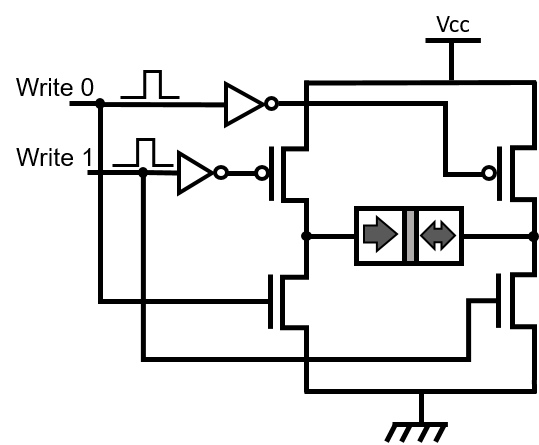}
	\caption{
    Programming circuit for magnetic tunnel junctions.}
	\label{fig:ProgModel}
\end{figure}

ST-MRAMs are based on magnetic tunnel junctions (MTJs), nanodevices composed of stack of various materials.
These materials implement two nanomagnets (reference and free magnet), separated by an oxide tunnel barrier.
The reference and free magnet may have either parallel or anti-parallel magnetizations, which constitutes the memory state of the device.
Due to the tunnel magnetoresistance (TMR) effect, a magnetic tunnel junction exhibits different resistance values in the parallel (\Rp) or antiparallel (\Rap) states. 
The TMR parameter is defined by:
\begin{equation}
\label{eq:TMR}
    \mathrm{TMR} = \frac{\Rap - \Rp}{\Rp}.
\end{equation}

Due to the spin transfer torque effect, the state of a magnetic tunnel junction can be switched by applying positive or negative electrical currents to it.
Unfortunately, the physics of spin torque switching is inherently stochastic  \cite{worledge_switching_2010, wang_bit_2012, vincent_analytical_2015, vincent2015spin,ranjan_approximate_2015, sampaio_approximation-aware_2015}, which is a source of intrinsic bit errors.
The physics of this phenomenon is now well understood.
As a reasonable approximation, spin torque switching may be modeled by Sun's model \cite{sun_spin-current_2000}.
The mean switching time is given by:
\begin{equation}
\label{eq:Sun}
    \tau = \tau_0 \frac{\Vc}{V - \Vc},
\end{equation}
where $\tau_0$ is a characteristic time, and \Vc{} is called the critical voltage of the junction. This equation is valid for voltages significantly higher than \Vc{}.

Many works have evaluated the statistical distribution of the actual switching time of the junctions. 
Here, we use the model proposed in \cite{vincent_analytical_2015}.
The distribution of switching time $t$ is given by  the gamma distribution:
\begin{equation}
\label{eq:Gamma_PDF}
	f_{\mathrm{sw}}(t;\,k, \theta) = \frac{t^{k -1} \exp{\left( -\frac{t}{\theta} \right)}}{\Gamma(k) \theta ^k },
\end{equation}
For the skewness $k$, we use the value suggested in \cite{vincent_analytical_2015} $k=16.0$ (corresponding to a relative standard deviation of $0.25$), and we take $\theta = \tau / k$.

We assume that the magnetic tunnel junctions are programmed with the standard circuit of Fig.~\ref{fig:ProgModel}, which allows applying positive or negative currents to the junction.
In this work, we propose to reduce the MTJ programming time in order to reduce the required programming energy, while increasing the BER.
This has been identified as the most efficient strategy to use MTJs as ``approximate memory'' in  \cite{locatelli2018use}, and here we look at the impact of this strategy on BNNs.

In all results, the circuit is simulated using the Cadence Spectre simulator, with the design kit of a \SI{28}{\nm} commercial technology.
The magnetic tunnel junction is modeled with a Verilog-A model described in \cite{locatelli2018use}, parametrized with values inspired by a \SI{32}{\nm} technology using perpendicular magnetization anisotropy \cite{khvalkovskiy_basic_2013, chun_scaling_2013}.
The diameter of the junctions is \SI{32}{\nm}, the thickness of the free layer is \SI{1.3}{\nm}, its
saturation magnetization is \SI{1.58}{\tesla}/\muo.
The resistance area (RA) product of the junction is \SI{4}{\ohm\um\squared}, its TMR value is $150\%$, and its critical voltage \Vc{} is \SI{190}{\mV}.
The junctions are programmed with a voltage of $2.0 \times \Vc$.
In Monte Carlo simulations, we consider mismatch and process variations of the transistors, as well as typical variations of MTJ parameters ($5\%$ relative standard deviation on TMR and \Rp{} \cite{worledge_switching_2010}).

\begin{figure}[ht]
	\centering
	\includegraphics[width=0.9 \linewidth]{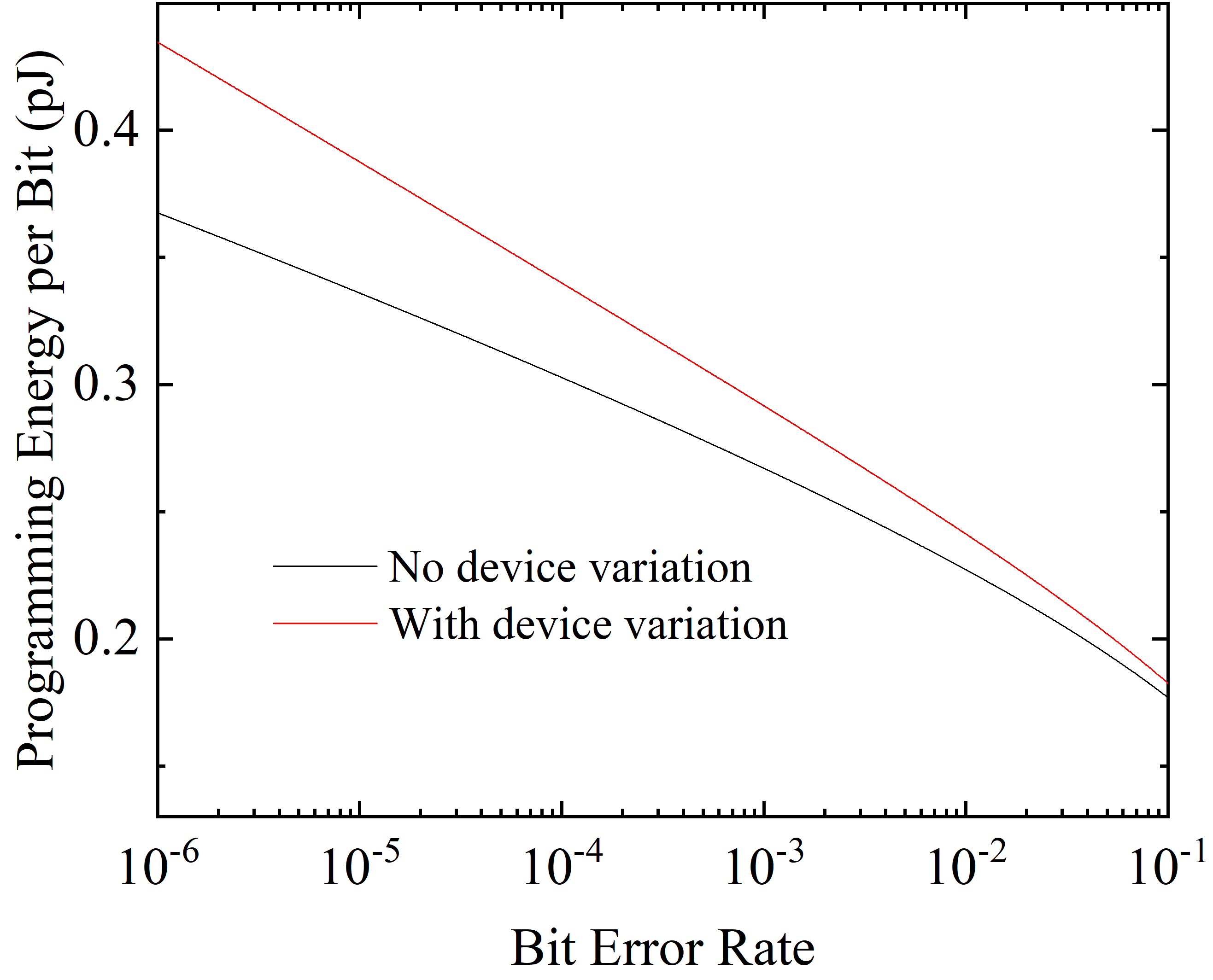}
	\caption{
    ST-MRAM programming energy per bit as a function of corresponding BER, ignoring (black line) or taking into account (red line) CMOS and MTJ device variations.
    Results obtained by Monte Carlo simulation.}
	\label{fig:BER}
\end{figure}

Fig.~\ref{fig:BER} presents the correspondence between BER and programming energy using the circuit of Fig.~\ref{fig:ProgModel}.
Two cases are evaluated:
considering only the intrinsic stochastic effects of the MTJs (black curve), and adding the effects of transistor and MTJ variability (red curve).
This curve confirms the existence of an interplay between programming energy and BER.

\begin{figure}[ht]
	\centering
	\includegraphics[width=0.9 \linewidth]{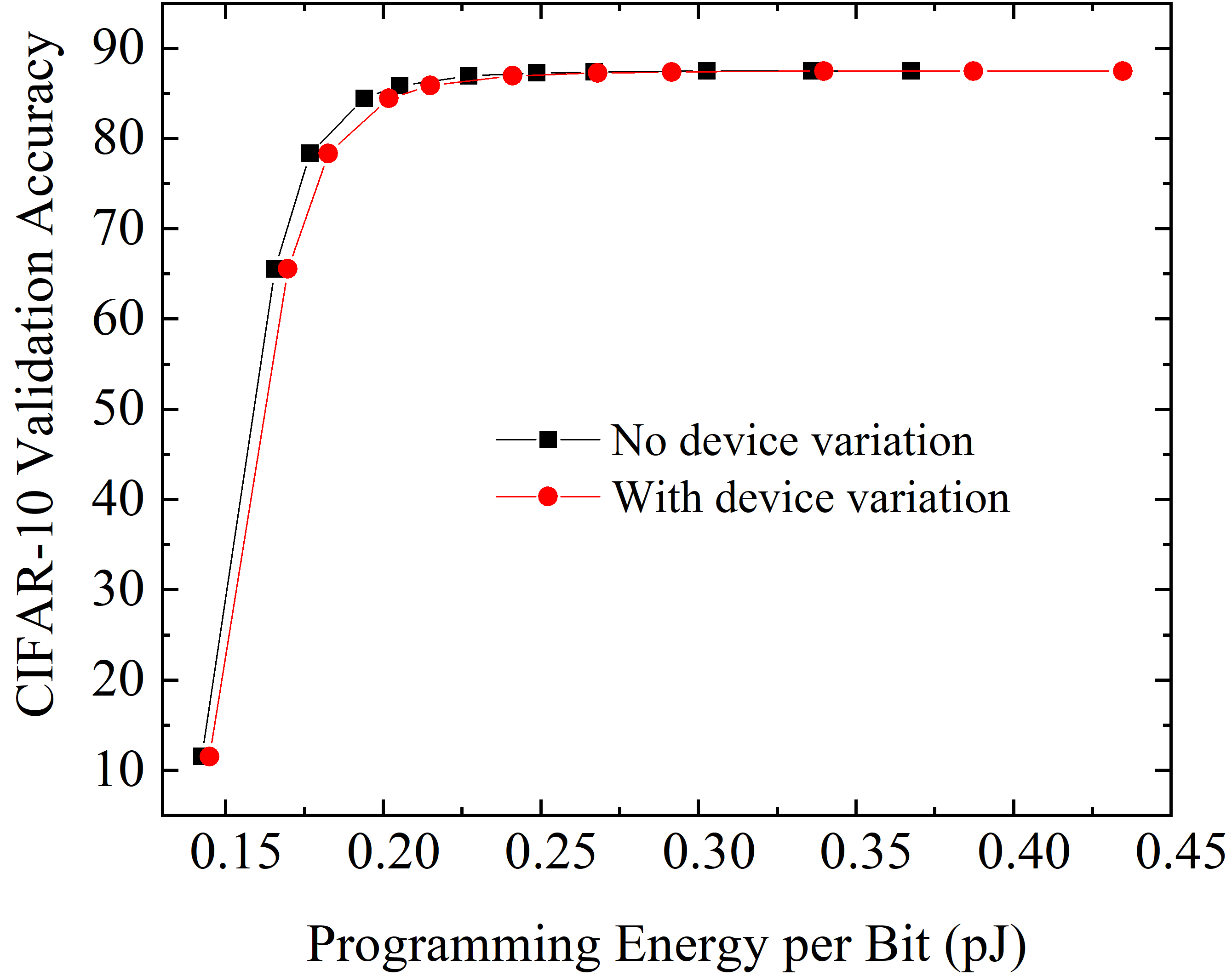}
	\caption{
     Validation accuracy of the convolutional binarized neural network trained on the CIFAR-10 dataset as a function of the ST-MRAM programming energy per bit, ignoring (black line) or taking into account (red  line) CMOS and MTJ device variations.}
	\label{fig:CIFAREnergy}
\end{figure}

\begin{figure}[ht]
	\centering
	\includegraphics[width=0.9 \linewidth]{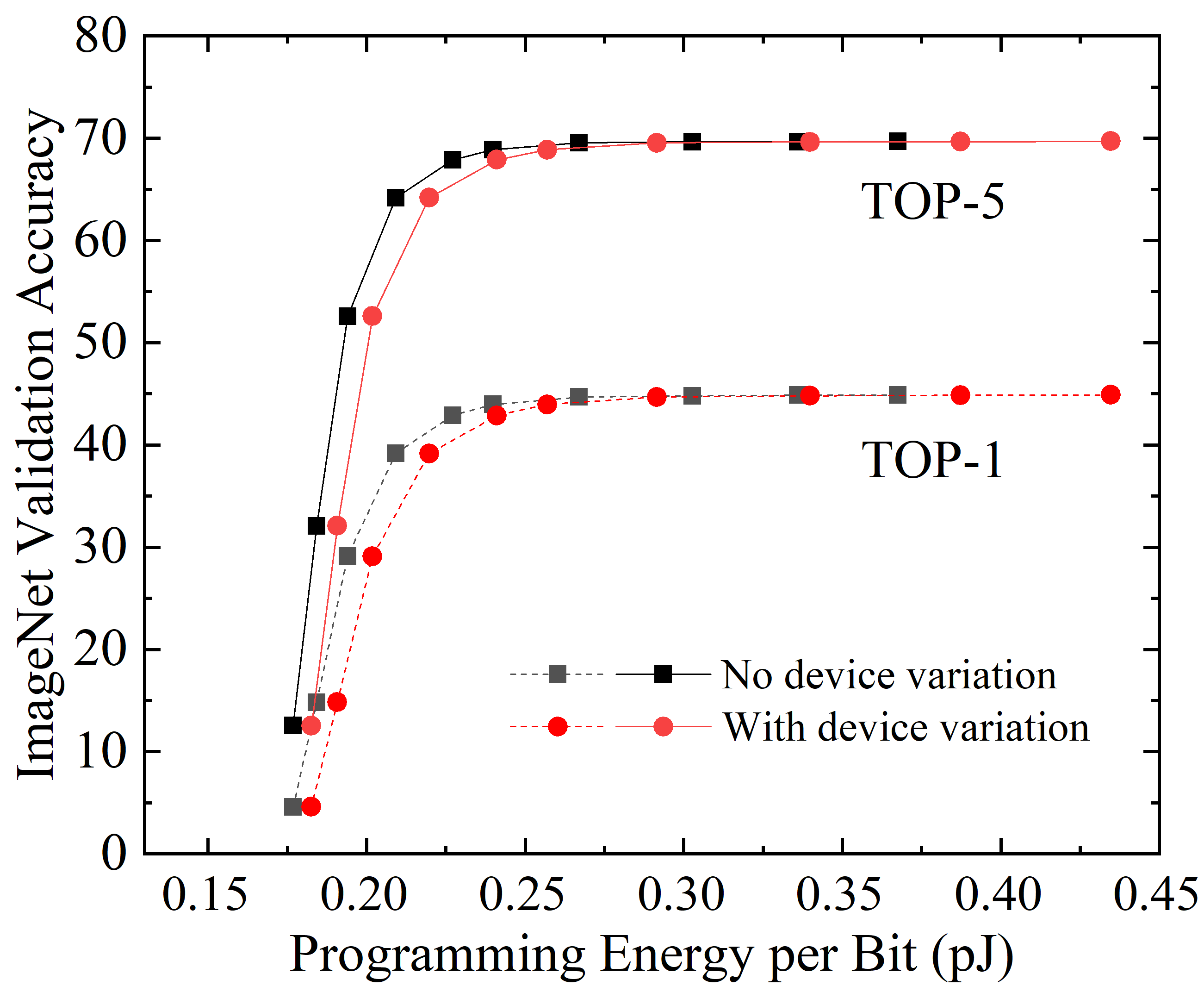}
	\caption{
    Top-1 and Top5 validation accuracy of binarized AlexNet trained on the ImageNet dataset as a function of the ST-MRAM programming energy per bit, ignoring (black line) or taking into account  (red  line) CMOS and MTJ device variations.}
	\label{fig:ImageNetEnergy}
\end{figure}

Figs.~\ref{fig:CIFAREnergy} and \ref{fig:ImageNetEnergy} associate the results of Figs.~\ref{fig:accuracy} and \ref{fig:BER} to highlight the interplay between programming energy and BNN accuracy for the CIFAR-10 task (Fig.~\ref{fig:CIFAREnergy}) and ImageNet (Fig.~\ref{fig:ImageNetEnergy}).
The points with highest programming energy correspond to a BER of $10^{-6}$.

We see that on both tasks the programming energy can be reduced approximately by a factor two with no impact on BNN accuracy.
This result is not only useful in terms of energy efficiency.
As the area of ST-MRAMs cells is dominated by transistor and not MTJs, this result means that we may be able to use lower area cells for BNNs (with lesser drive current) than for conventional memory applications. 
This strategy of reducing programming energy if one accepts increased BER is also applicable to other memory technology such as oxide-based RRAM \cite{hirtzlin2019outstanding}, even if the underlying physics of these devices differ considerably from ST-MRAMs.

\section{Conclusion}

In this work, we saw that Binarized Neural Networks can tolerate high bit error rates (up to $10^{-3}$) on their binarized weights, even on difficult tasks such as ImageNet.
This makes them especially appropriate for in-memory hardware implementation with emerging memories such as ST-MRAMs.
Not only can we use ST-MRAMs without the use explicit error correcting codes, but we can also program ST-MRAMs with reduced energy.
Based on physical modeling of ST-MRAMs, we showed that a factor two in programming energy can be saved.

These results highlight that neural networks differ from more traditional forms of computation.
In a way that is reminiscent of brains, which function with imperfect and unreliable basic devices, perfect device reliability might not be necessary for hardware neural networks.

\section*{Acknowledgment}
This work was supported by the Agence Nationale de la Recherche grant NEURONIC (ANR-18-CE24-0009) and the European Research Council Grant NANOINFER (715872).

\bibliography{IEEEabrv,Mabibliotheque}
\bibliographystyle{IEEEtran}

\end{document}